# Experimental evidence for topological phases in the magnetoconductance of 2DEG-based hybrid junctions

Kaveh Delfanazari[1,2,3,*], Llorenç Serra[4], Pengcheng Ma[2], Reuben K. Puddy[2], Teng Yi[2], Moda Cao[2], Yilmaz Gul[2], Ian Farrer[2,5], David A. Ritchie[2], Hannah J. Joyce[1], Michael J. Kelly[1,2], Charles G. Smith[2]

[1]Electrical Engineering Division, Engineering Department, University of Cambridge, Cambridge CB3 0FA, UK
[2]Department of Physics, Cavendish Laboratory, University of Cambridge, Cambridge CB3 0HE, UK
[3]James Watt School of Engineering, University of Glasgow, Glasgow G12 8QQ, UK
[4]IFISC (UIB-CSIC) and Physics Department, University of the Balearic Islands, E-07122 Palma, Spain
[5]Department of Electronic and Electrical Engineering, University of Sheffield, Mappin Street, Sheffield, S1 3JD, UK
*Corresponding author email: kd398@cam.ac.uk

Dated: 28072020

**While the application of *out-of-plane* magnetic fields was, so far, believed to be detrimental for the formation of Majorana phases in artificially engineered hybrid superconducting-semiconducting junctions, several recent theoretical studies have found it indeed useful in establishing such topological phases [1-5]. Majorana phases emerge as quantized plateaus in the magnetoconductance of the hybrid junctions based on two-dimensional electron gases (2DEG) under fully *out-of-plane* magnetic fields. The large transverse Rashba spin-orbit interaction in 2DEG, together with a strong magneto-orbital effect, yield topological phase transitions to nontrivial phases hosting Majorana modes. Such Majorana modes are formed at the ends of 2DEG-based wires with a hybrid superconductor-semiconductor integrity. Here, we report on the experimental observation of such topological phases in Josephson junctions, based on In$_{0.75}$Ga$_{0.25}$As 2DEG, by sweeping *out-of-plane* magnetic fields of as small as $0 < B_\perp (\text{mT}) < 100$ and probing the conductance to highlight the characteristic quantized magnetoconductance plateaus. Our approaches towards (i) creation and detection of topological phases in small out-of-plane magnetic fields, and (ii) integration of an array of topological Josephson junctions on a single chip **pave the ways for the development of scalable quantum integrated circuits for their potential applications in fault-tolerant quantum processing and computing.**

Topological superconductors with a superconducting gap in their bulk and topologically protected metallic states on their edges and surfaces host exotic quasiparticles such as anyons and Majorana fermions. Such exotic (non-abelian) quasiparticles have been proposed for manipulation and application in fault-tolerant topological quantum processing and computing because their quantum states defined by a pair of Majorana zero modes (MZMs) work as a nonlocal qubit [1-18]. Therefore, the search effort for the realization of topological superconductors, in both artificial and intrinsic forms, has been considerably increased recently.

Artificially engineered topological superconductors in hybrid superconductor-semiconductor (S-Sm) material platforms have attracted substantial attention because they offer stimulating opportunities to create and manipulate non-Abelian MZMs [10-13]. In most studies of hybrid junctions based on one-dimensional (1D) wires or quasi-1D systems, the appearance of the zero-bias peak (ZBP) in the DC spectroscopy of the differential conductance above a relatively high critical in-plane magnetic field (usually $\geq 0.5$ T) has been interpreted as evidence for the MZMs [6,8,17,18]. However, (i) the application of such a relatively large in-plane global magnetic field to achieve the topological phase transitions makes such a hybrid system unfeasible in technological applications, and (ii) the ZBP can be a debatable argument, especially when it comes to zero magnetic fields. It will be awkward to distinguish them from non-Majorana origin ZBP, such as Kondo, Josephson effect, and quantum dot excitation. Recent theoretical and experimental investigations also suggest that the ZBPs can appear in the presence of disorder even when the junction is in the non-topological phase, so with the non-Majorana origin, and emerge due to ordinary Andreev bound states that exist close to zero energy [8]. Such Andreev states will have a similar dependence on the in-plane magnetic fields as the Majorana modes.

Consequently, the experimentally observed appearance and disappearance of the ZBP as a function of in-plane magnetic fields do not represent unambiguous evidence for the Majorana modes [1-5,8]. Therefore, a different approach towards the realization of topological superconductivity that does not rely on the ZBP is sought [1-5,7,8,19,20]. An alternative experiment towards the realization of topological superconductivity is the 4π periodic Josephson effect that can be observed by microwave spectroscopy [7]. The 4π-periodic Josephson effect was believed to represent a more reliable signature of Majorana fermions than the tunneling spectroscopy of the ZBP. Recently, however, it was shown that the 4π-periodic Josephson effect could occur under certain conditions even for Josephson junctions made of trivial superconductors with no MZMs. Thus, the aforementioned experimental results weigh in favour of Majorana interpretation but do not represent definitive proof [8].

In this paper, building on the idea of exploring the unambiguous detection of MZMs in hybrid systems, we demonstrate a novel approach that emerges as quantized conductance plateaus and doesn't rely on the debated arguments above. We report on the first experimental observation of topological phases in hybrid junctions, on a two-dimensional electron gas (2DEG) platform, in a fully perpendicular (out-of-plane) magnetic field. The advantages of our innovative approach include: (i) the perpendicular magnetic field has a paramount influence on the motion of quasiparticles in the plane; the scenario of the well-studied quantum Hall effect. (ii) the perpendicular magnetic field geometry is also more convenient experimentally since, in this case, the field has a maximal influence, and thus lower fields are more effective for phase tuning in a topological system. (iii) the strong magneto-orbital effect and transverse Rashba spin-orbit in 2DEG leads to nontrivial Majorana phases for relatively low values of the perpendicular fields in our 2D junctions. (iv) the novel Majorana architecture based on Josephson junction array on 2DEG systems, demonstrated in this paper would allow making a complex network of such devices, so addressing a significant issue of the scalability of topological quantum devices, for the first time, comparing with nanowires as they can't be used in the real quantum world due to their 1D geometry limitation. (v) the emergence of topological superconductivity in 2D Josephson junctions, as the building block of Majorana qubits, is entirely controlled by the external out-of-plane magnetic fields, which offers better control over the topologically nontrivial phases and the probability to observe the Majorana phases at much weaker magnetic fields [1-5,16]. (vi) and finally, our proposed 2DEG platform doesn't rely on the fine-tuning of device parameters, e.g. chemical potential, as one essentially needs for a 1D nanowire platform [16-18].

We show that transitions to topological phases can be observed as signatures in the junctions' magnetoconductance. We furthermore demonstrate that such phases can be tuned with very small, out-of-plane magnetic fields, a magnetic field ranges of zero to 100 mT. Our observation is based on the recent theoretical proposal for the existence of topological phases in the magnetoconductance of hybrid junctions based on a 2DEG coupled to superconducting contacts [1]. The single (NS) and double (NSN) hybrid junctions reveal topological phase transitions to nontrivial phases as a consequence of an enhanced magneto orbital effect, due to applied external magnetic fields, and in the presence of strong transverse Rashba spin-orbit (SO) interaction in a 2DEG. Such junctions host Majorana modes in their hybrid segment. The presence of a potential barrier at the junction interface reveals the Majorana phases as quantized plateaus of high conductance for narrow junctions. In wider junctions, where the SO interaction becomes stronger relative to the transverse confinement, the sequence of multiple transitions trivial-topological- trivial, and so on, takes place at low fields with characteristic magnetoconductance features. The conductance yields quantized plateaus in the topological phases with a Majorana mode, while it is quenched in the trivial regions. These differences are caused by the presence of an interface barrier that strongly suppresses the Andreev reflection [21] and reduces the conductance of the trivial regions.

On the contrary, the Majorana state of the nontrivial regions allows maintaining a high conductance even in the presence of an interface barrier. Therefore, the influence of the interface barrier in trivial and topological regions is very different. Importantly, the quenched conductance of the trivial regions causes a dip in conductance, followed by an enhancement when the field is such that a nontrivial phase is entered.

Figure 1 (a) shows the schematic view of the studied Josephson junctions in the presence of an out-of-plane magnetic field. The superconducting Nb leads (grey parts on two sides of the device) act on the piece of the 2DEG beneath and between them to form the hybrid 2DEG (red regions) by the proximity effect. The quantum transport then occurs between the top Nb (grey)- hybrid 2DEG (red)- normal 2DEG (black)- hybrid 2DEG (red)- bottom Nb (grey) in the presence of out-of-plane magnetic fields, see Fig. 1 (a).

Note that when the vertical field is switched off, the normal 2DEG (black) no longer exists between two Nb pads and the device gets the form of Nb- hybrid 2DEG-Nb Josephson junction because the Andreev reflections in such ballistic junctions carry supercurrent in the normal 2DEG part (see temperature/magnetic field dependent supercurrent data in SI). Figure 1 (b-c) is the result of a theoretical model for our Josephson junctions with two superconducting contacts based on Ref. [1] for an illustrative case. The phase boundaries in the $\mu$-$B$ plane correspond to the condition of a $k = 0$ propagating modes at zero energy of the hybrid junction. That is the $k = 0$ gap-closing lines in the $\mu$-$B$ plane. Figure 1 (b) shows the phase boundaries for a junction of width $w = 1$ µm. Note that there are many boundaries packed at low field. If one increases the field along the cyan dotted lines, the topological junction changes phase as 0- 1- 0- 1… Majorana modes each time a topological phase line is crossed. That is, the junction is evolving along the sequence of trivial- topological- trivial- topological... phases [1]. As the field is increased, the number of propagating modes of the normal section decreases. For instance, along the cyan dotted lines of Fig. 1 (b), the evolution each time a line is crossed is $N$- ($N$-1)- ($N$- 2) - … , with $N$ being the number of modes at zero fields. The Andreev reflection is perfect for all channels when the topological section is in a Majorana mode phase, and that is imperfect in phases with no Majorana modes. The Andreev reflection of the trivial phases decreases quadratically as the field is increased from the initial value of that phase.

Figure 1 (c) shows the modeled conductance for the field sequence along three-dotted cyan lines in the left panel (Fig. 1b). Note that the quadratic decrease of the trivial phases leads to dips before the onset of the next phase. These dips are especially apparent for the initial transitions where the number of propagating modes is higher. All three conductance curves show flat quantized plateaus in steps of around $2e^2/h$. The schematic side view of the studied 2D junctions is shown in Fig. 1 (d). The devices are fabricated by contacting Nb (shown in grey) to 30 nm thick, high-quality $In_{0.75}Ga_{0.25}As$ two-dimensional electron gas that is formed 120 nm below the surface of an $In_{0.75}Al_{0.25}As$/GaAs heterostructure wafer [23-26]. Figure 1 (e) is the scanning electron micrograph (SEM) of the quantum chip with an array of 2D junctions, eight independently controlled Josephson junctions on this chip. In our CAD design, two Nb leads of width $w$= 4 μm are separated with a length $L$= 850 nm through $In_{0.75}Ga_{0.25}As$ quantum wells. However, these dimensions are subject to change after fabrication by wet-etching technique and especially under the application of vertical fields, as we explain below [23]. The SEM of one junction is shown in Fig. 1 (f). The motivation behind our approach with quantum integrated circuits includes fabricating and measuring an array of devices in one fridge cool-down with minimum experimental condition variations. From an array of eight devices on a single chip, one failed during the fabrication process. We observed quantum transport in three devices, but with weaker induced superconductivity and high field conductance structures- they will be discussed in detail elsewhere.

The other four devices show strongly induced superconductivity, multiple Andreev reflections, and large supercurrent of up to 2 µA [see SI file]. These junctions are named as D1-4 and discussed in full detail in this paper. The elastic mean free path of these junctions is around $l_e = e^{-1}\hbar\mu_e\sqrt{2\pi n_s} \cong 2\ \mu m$ thus all junctions are in the ballistic regime [23] and the coherence length of around 2 µm in 2DEG calculated from Equation $\xi_N = \hbar v_{FN}/\pi\Delta_{Nb}$ suggest that they are in a short junction limit, too [23]. An average normal state resistance of $R_N$ = 0.2 kΩ is also measured in these junctions.

Figure 2 (a) shows the quantized conductance observation as a function of the out-of-plane magnetic field for D1 measured at $T$= 50 mK and 620 mK. It shows the magnetic field sweep for two different directions: the red and orange curves show a sweep from +0.4 T to -0.4 T. In contrast, the blue and dark green curves show the reversed sweep from -0.4 T to +0.4 T. By increasing field from 0 T to +/-0.4 T, the conductance drops and quantized plateaus appear when the magnetic field is between 0 and +/-0.08 T, and the base temperatures range is between $T$= 50 mK and 320 mK. Figure 2 (b) is a close look of Fig. 2 (a) at small magnetic field ranges to highlight the plateaus for both directions of magnetic fields. By reversing fields from +/-0.4 T to 0 T, the conductance of the junction increases. Between $B$= +/-0.08 T and 0 T (see Figs 2 (a-d)), the conductance plateaus are again observed for temperatures below $T$= 320 mK. The plateaus are slightly different for two magnetic field sweeps directions: from -0.08 T to 0 T and from 0 T to +0.08 T, as shown in Fig. 2 (e) and (f), respectively. We believe that the difference might be due to the occurrence of bunching of transitions in low fields, and we may only probe one feature due to experimental resolution limitation. The heating effects in the device for two negative/positive magnetic field sweep directions might also affect the conductance plateaus. We can see that the conductance plateaus with quantized values in steps of around 2 $e^2/h$ observed in the experiment are in excellent agreement with our model which predicts a roughly similar amount of transport modes with quantized conductance in hybrid 2D junctions (see Fig. 1 c and the discussion in materials and methods section about the model).

The conductance curve at $T=$ 50 mK shows a sequence of low field phase transitions on the junction conductance. It suggests that a scattering barrier may be formed at the interface of the junction that strongly quenches the conductance of the trivial phases, causing a dip in the conductance curve. In contrast, the conductance remains quantized in topological ones due to the enhanced Andreev reflection. To further investigate the origin of the observed conductance plateaus, we look at the effect of the temperature and magnetic field sweep rate. The quantized conductance at temperatures between $T=$ 50 mK and 320 mK for magnetic fields between $-0.08 < B$ (T) $< 0$ and magnetic fields between $0 < B$ (T) $< +0.08$ branches of $G$ for field sweeps from -0.4 T to +0.4 T are replotted in Figs. 2 (e) and 2 (f), respectively. The corresponding number of plateaus as a function of magnetic field and temperature is shown in Fig. 3 (a).

We observe five plateaus at the lowest base temperature ($T=$ 50 mK) in which this number decreases as temperature increases, due to the reduction in the Andreev reflection probabilities. Still, their positions (plateaus steps) are almost unchanged. Figure 3 (b) shows the quantized conductance for D2 observed for reversed applied magnetic fields ranging of $0 < B_\perp$ (mT) $< +80$ at $T=$ 50 mK at different field sweep rates: 0.1 T/hrs (orange), 0.2 T/hrs (blue), 0.5 T/hrs (red) and 1 T/hrs (black). As can be inferred from this figure, the number of plateaus and their widths do not change, which suggests that the phase transition is also independent of the field sweep rate. We, therefore, exclude any thermal effects.

Figure 4 shows the normalized conductance as a function of the applied out-of-plane field at temperature $T=$ 50 mK for D3 at 1$^{st}$ run for sweep direction from left to right (a), 2$^{nd}$ run for sweep direction from right to left (b), 3$^{rd}$ and 4$^{th}$ runs for sweep direction from left to right (c) and right to left (d). Looking at Fig. 4, it is remarkable that the conductance of the hybrid junction in trivial phases is quenched by showing robust dips, but it is unaffected in topological regions.

Figure 5 shows quantized plateaus for D4 measured at temperatures $T=$ 50 mK (a), 120 mK (b), 220 mK (c), and 320 mK (d), for a narrower range of magnetic fields with sweep direction from +0.4 T to 0 T. Here, we highlight the observed quantized plateaus with apparent dips that are a manifestation of the topological phase transitions of the hybrid junction. All our investigations show that the conductance plateaus observation is quite robust and reproducible in different devices at different temperatures, magnetic field sweeps rates, and also magnetic field sweep directions. There are small variations in the dip and peak positions of the plateaus of different devices. They may be explained in a way that the effective dimensions of the junctions after fabrication, by a wet etching method, might differ from the design values giving rise to a different number of active modes in the junctions, thus affecting their conductance.

From the given values of $R_N \approx 0.2$ k$\Omega$, electron density $n_s$, mobility $\mu_e$ and interface transparency of our high-quality Nb-In$_{0.75}$Ga$_{0.25}$As-Nb junctions [23] we can estimate the number of modes to be around $N \cong 100$ in a junction with designed width $w=$ 4 μm [18]. However, our junctions are made by the wet-etch technique, so the effective dimension of the junctions would normally be subject to changes after fabrication, and this can be estimated from the number of modes in the conductance of the junctions [17,18]. From the conductance of our devices in the experiment, we can infer the number of transport modes to be roughly 50 $\leq N \leq$ 80. This fits roughly in a junction width of around 0.8 $\leq w$ (μm) $\leq$ 2. By looking at our calculated conductance in Fig. S5, Fig. 1, and Fig. S6 for Josephson junctions of widths $w=$ 0.75 μm, 1 μm, and 1.25 μm, respectively, we find a good agreement between experiment and theory both in the number of modes and also in the observed quantized conductance plateaus with the steps of roughly around $2\ e^2/h$. A similar change in the effective dimensions of Josephson junctions was also reported quite widely in Josephson junctions of different material systems, for instance, in 2DEG [18, 27] as well as in nanowire [28].

In short, we presented the evidence for topological phases in the magnetoconductance of hybrid junctions in a scalable quantum circuitry under a fully vertical (out-of-plane) magnetic field. Our experimental results are consistent with the topological phase transitions in the 2D material system in small out-of-plane magnetic fields. We introduced a radically different but quite efficient and robust approach towards topological phase detection in planar Josephson junctions. We demonstrated that array of topological junctions could be integrated on a single chip so one can fabricate array and network of 2D topological junctions on a single chip for various purposes such as for braiding the Majoranas or measuring/assessing their reproducibility, quantum yield, and reliability for their use in quantum technologies in one fridge cool-down with minimum experimental ambient condition variations. Our approach may also help the development of robust quantum integrated circuitry for applications in decoherence-free quantum computation and secure communication.

**Materials and Methods**

The $In_{0.75}GaAs/In_{0.75}AlAs/GaAs$ quantum well was grown by molecular beam epitaxy (MBE) [23-26, 29]. The 30 nm thick 2DEG InGaAs quantum well with density $n_s=2.24\times10^{11}$ (cm$^{-2}$) and mobility $\mu_e=2.5\times10^5$ (cm$^2$/Vs) in the dark was formed 120 nm below the wafer surface. The fabrication methods of various types of 2D Josephson junctions on a semiconducting chip have been extensively discussed in our previous works [23-26]. Here, we bring a summary:

Using photolithography and wet etching, in which a sulfuric acid solution of compositions $H_2SO_4$, $H_2O_2$, and $H_2O$ were used, we created an active region (a raised area referred to as mesa structure and shown by two parallel dashed line in Fig. 1 e) with length l′ = 1440 μm and w′ = 160 μm in the middle of our chip where all the eight identical junctions are patterned and fabricated (see Fig. 1). To make junctions, after photolithography patterning, we removed the top InGaAs and InAlAs layers in the patterned area using wet etching. The etch produces a 120–140 nm deep trench in this area, which is around the 120 nm depth of the $In_{0.75}Ga_{0.25}As$ quantum well. This was followed by the deposition of a ≈130 nm Nb film (about three times larger than the London penetration depth of Nb, which is 40 nm) to make high-quality superconducting contacts to the 2DEG, using DC magnetron sputtering in an Ar plasma. According to the design, each junction has a length of $L$ = 850 nm at the shortest path, which increases to 26 μm at the edge of the active region and a width of $w$ = 4 μm as shown in Figure 1. The ohmic contacts were made of gold/germanium/nickel (AuGeNi) to get a low resistance and good chemical bond (adhesion) to the semiconductor substrate and placed 100 μm away from two Nb– $In_{0.75}Ga_{0.25}As$ quantum well interfaces to reduce any influence of the normal electrons on the interface. Not all of the ohmic pads were used for source-drain bias measurements in this study. Gold wire of diameter 20 μm was used for the bond wires. The quantum transport measurements of conductance as a function of temperature and magnetic fields were performed by using a standard lock-in technique by superimposing a small ac-signal at a frequency of 70 Hz and an amplitude of 5 μV to the junction dc bias voltage and measuring the ac-current.

**Numerical modeling:**

Our model is based on the recent theoretical proposal on evidence for Majorana phases in the magnetoconductance of topological junctions based on 2DEGs [1].

A model Hamiltonian of a hybrid junction with a width $w$, where $-w/2 < y < w/2$, can be written as

$$H = \left(\frac{p_x^2 + p_y^2}{2m} - \mu\right)\tau_z + \frac{\alpha}{\hbar}(p_x\sigma_y - p_y\sigma_x)\tau_z$$
$$+ \Delta_B\sigma_z + \Delta_0\tau_x + \frac{\hbar^2}{2ml_z^4}y^2\tau_z - \frac{\hbar}{ml_z^2}yp_x - \frac{\alpha}{l_z^2}y\sigma_z, \quad (1)$$

Here, $\alpha$, $\Delta_B$, $\Delta_0$, $\mu$, $m$, and $\sigma$ ($\tau$) is the SO interaction, Zeeman, pairing parameters, chemical potential, effective mass, and Pauli matrices in spin (particle-hole) space, respectively. The Zeeman energy $\Delta_B$ is related to the field $B$ by $\Delta_B = g^*\mu_B B/2$ with $g^* = 15$ as the gyromagnetic factor. The other parameters of the model are induced superconductivity $\Delta_0 = 60$ μeV (close to $\Delta_0 \cong 100$ μeV observed in the experiment for most junctions fabricated in our group [23]), the strength of the Rashba spin-orbit interaction $\alpha = 20$ meVnm [18], and $m = 0.039$ $m_e$ [31]. The orbital field terms, the last three terms in Equation (1), are related to the magnetic length $l_z^{-2} = eB/\hbar c$. We highlight here that the last three terms in Equation (1) are often neglected in studies of Majorana modes in hybrid junctions of almost 1D wires. By performing the microscopic modeling of narrow 2D single SN junctions, the conductance $G$ can be calculated from [1]

$$G = \frac{e^2}{h}(N - R + R_A) \quad (2)$$

and in a double NSN junction, in the same way, the conductance gets the form of

$$G = \frac{e^2}{h}(T + R_A) \quad (3)$$

where, $N$, $R$, $R_A$, and $T$ are the number of incident electron modes, normal reflection probability, Andreev reflection probability, and transmission from the left to right normal contact, respectively.

When the junction width increases, the microscopic solution of the scattering problem, especially for SNS double junctions, becomes complicated.

The difficulties stem from a large number of modes within a reduced energy interval and to the strong fluctuations with the magnetic field. However, Hoppe *et. al* demonstrated that the conductance of a long SNS double junction $G_{SNS}$ is given by the Andreev reflection $R_A$ of a single SN junction so that the SNS junction conductance can be written as $G_{SNS}= R_A e^2/h$ [22]. In phenomenological modeling based on Ref. [1], the magnetoconductance $G(B)$ of an SNS junction for fields $B \in [B_1, B_2]$ can be written as

$$G(B) = \frac{e^2}{h} \begin{cases} N_i & \text{if topological} \\ N_i \left[1 - \gamma \left(\frac{B - B_1}{B_2 - B_1}\right)^2\right] & \text{if trivial} \end{cases} \quad (4)$$

where, $B_1$ and $B_2$ indicate initial and final fields of a given phase of the hybrid S-Sm 2DEG strip, $\gamma$ is a phenomenological parameter and $N_i$ is the number of incident electronic modes from the normal side of the junction. As anticipated [1], a perfect Andreev reflection in topological phases is assumed and a parabolic decrease in trivial phases.

From data shown in Figs. 2, in addition to the plateaus observation, it is also found that the conductance curve peaks *asymmetrically* near $B= 0$, and hysteresis is observed when the applied fields are reversely swept. This suggests that the internal effective magnetic field $B_{eff}$ *slightly* differs from the applied external out-of-plane field $B_\perp$, maybe due to tiny magnetic interactions in our devices (see the conductance maxima in Fig. 2 (e) and Fig. 4 that peaks not at $B= 0$ but at slightly higher fields). A tiny magnetic impurity may stay in the sample holder or in the chip, or in the evaporation/sputtering chambers where the ohmic contacts have been deposited/sputtered [23]. Semiconducting wafers with a significantly reduced number of defects, and therefore with high electron mobility, can be produced by the molecular beam epitaxy, but the defects cannot be avoided entirely in the device [30]. See Fig. S7 and the corresponding explanation for a numerical model of this effect. A full analysis of this effect is out of the focus of the present work and will be discussed in detail elsewhere.

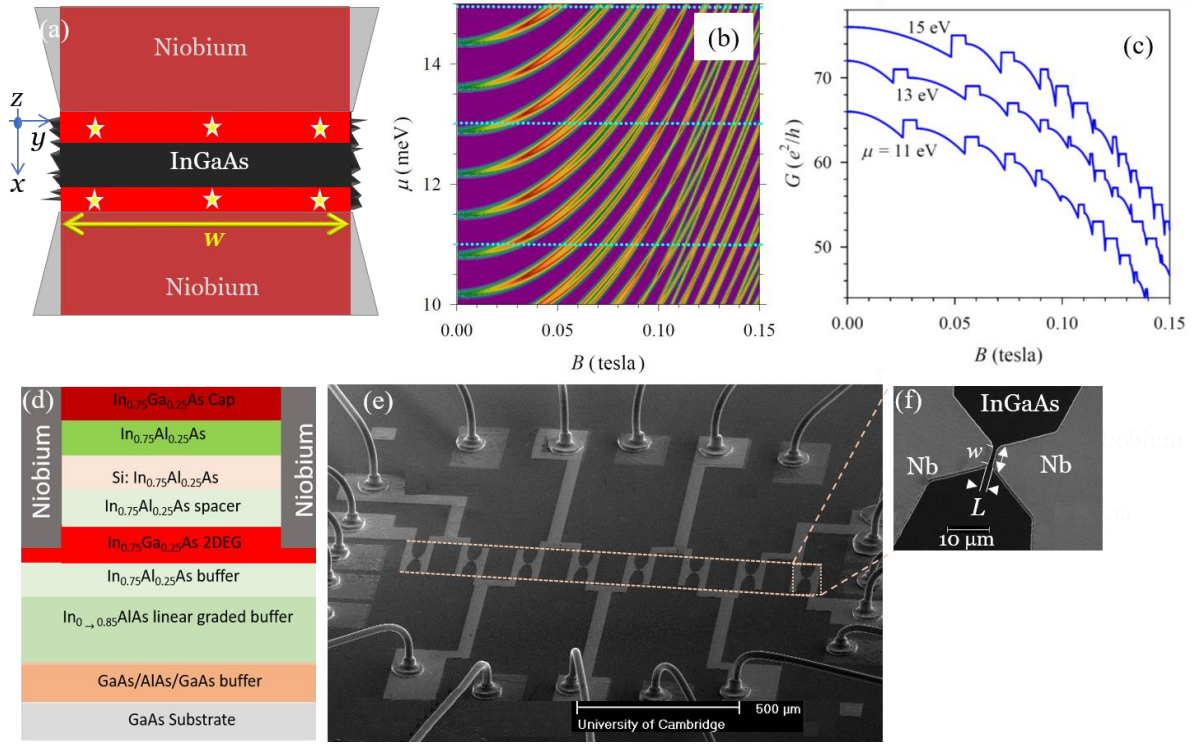

Figure 1. **Topological phases in hybrid Josephson junctions:** (a) A schematic view of a hybrid Josephson junction, under vertical magnetic fields. The superconducting Nb leads (grey parts) act on the piece of the 2DEG beneath and between them to form the hybrid 2DEG (red parts) by the proximity effect. The quantum transport proceeds between the top Nb (grey)- hybrid 2DEG (red)-normal 2DEG (black)- hybrid 2DEG (red)- bottom Nb (grey) in the presence of vertical magnetic fields. Stars are the boundary between normal and hybrid 2DEGs where Majorana modes are expected. Note that when the vertical field is switched off, the normal 2DEG (black) no longer exists between two Nb pads and the device gets the form of Nb- hybrid 2DEG-Nb Josephson junction because the Andreev reflections in the ballistic junctions carry supercurrent in the normal 2DEG part (see text and SI figures). (b) Phase diagram of the hybrid 2DEG (red in Fig. 1a) showing the boundaries between trivial (magenta) and topological regions with width $w$= 1 μm, a value close to the effective width of the junction in the experiment. The phase boundaries of the diagram have been calculated as the $k = 0$ gap closings. We consider superconductivity $\Delta_0$= 60 μeV in hybrid 2DEG which is close to the induced gap value observed in experiment. (c) Theoretical conductance of the junction taking the sequence of magnetic fields for three line cuts as cyan dotted lines in panel (b). The model assumes a perfect Andreev reflection in the topological phases (plateaus), and a decreasing Andreev reflection AR (dips) with increasing field in the trivial phases due to interface scattering. (d) The schematic view of an In$_{0.75}$Ga$_{0.25}$As/In$_{0.75}$Al$_{0.25}$As/GaAs heterostructure. Superconducting Nb leads are attached to high quality In$_{0.75}$Ga$_{0.25}$As quantum wells to form hybrid 2DEG by proximity induced superconductivity. (e) Scanning electron microscope image of the hybrid quantum circuit with an array of 2D symmetric and planar junctions on a chip. The superconducting parts are niobium (Nb) and shown in grey. The ohmic contacts are etched down to 2DEG region, for the purpose of tunneling measurements between Nb and 2DEG. The area between Nb and ohmic pads are all etched away except the active area (mesa shown with two parallel dashed-brown lines) where Josephson junctions are formed (dashed-brown square for one junction). (f) An SEM image of one junction on the chip.

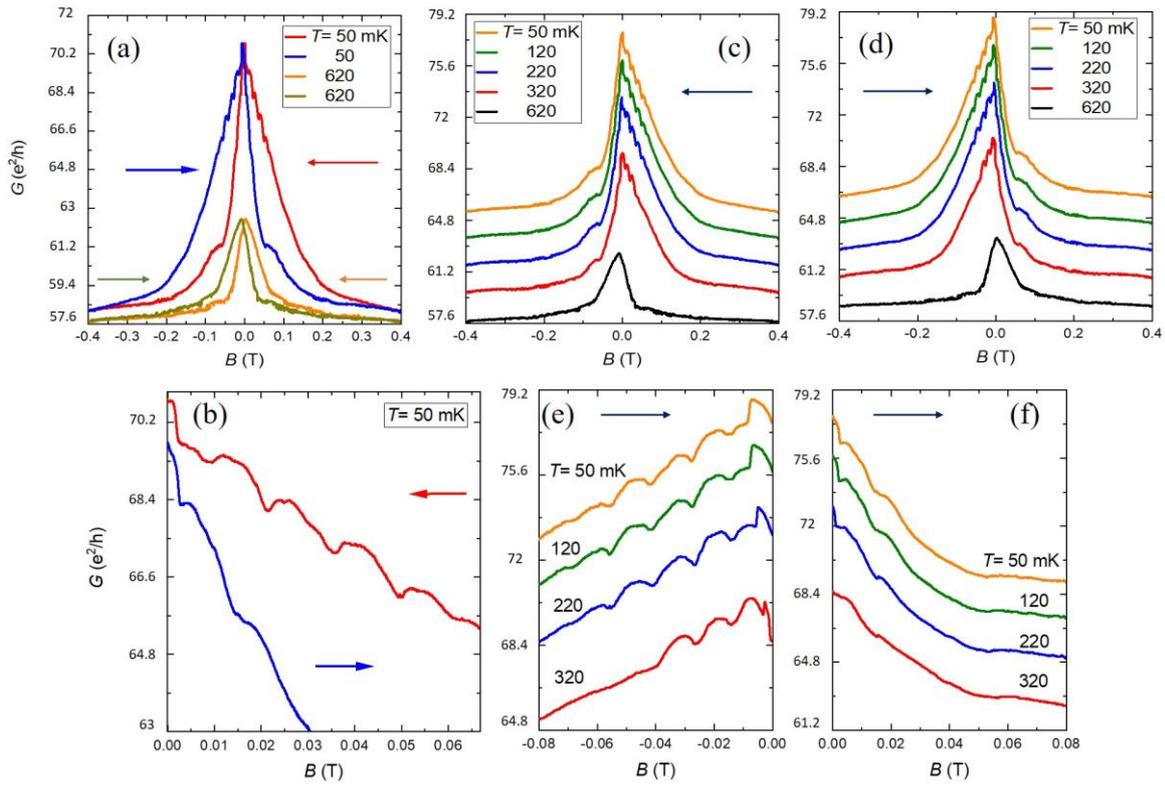

Figure 2. **Temperature dependence of the quantized conductance in topological Josephson junction:** (a) conductance vs. applied out-of-plane magnetic field $B_\perp$, at $V_{SD}= 0$ and at different base temperatures, for D1. (b) same information for a smaller range of magnetic fields to highlight that the quantized plateaus are observed in both directions. quantized conductance for magnetic field $B_\perp$ sweep directions from +0.4 T to -0.4 T (c) and from -0.4 T to +0.4 T (d). Strong asymmetric conductance $G$ curves are observed. (e)-(f) data shown in (d) are replotted for narrower magnetic field $B_\perp$ ranges to highlight the plateaus. The plateaus are a bit broadened for the increasing fields (possibly due to coincidence of bunch of transitions at low $B_\perp$ (as predicted in the model) and one has been probed in the experiment. (e) all curves in Figs. (c)-(f) are shifted vertically by 0.9 $e^2/h$ for clarity.

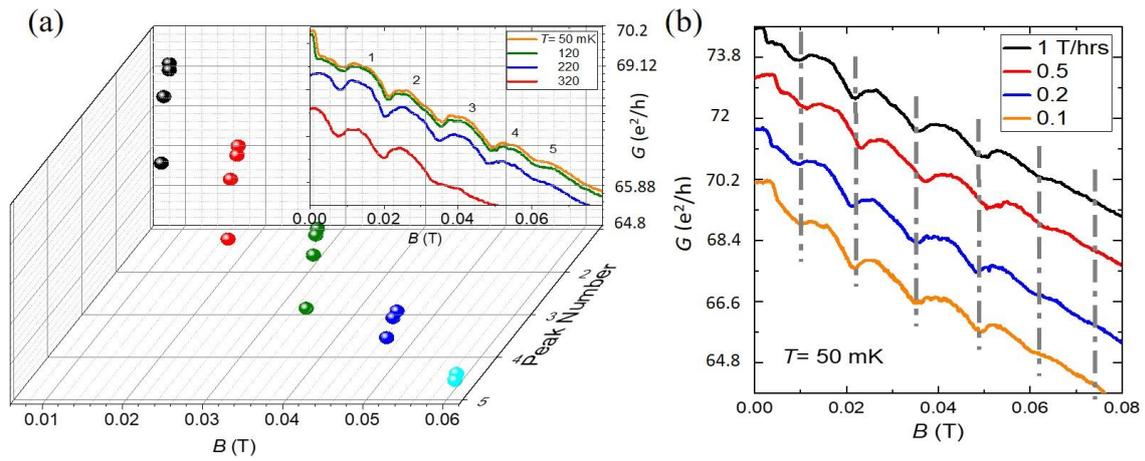

Figure 3. **Temperature and magnetic field sweep rate dependences of the quantized conductance in topological Josephson junctions:** (a) plateaus observed in D1 are numbered (inset) and plotted as a function of temperature and magnetic field $B_\perp$. (b) quantized conductance observed for reversed applied magnetic fields ranging of $0 < B_\perp$ (mT) $< 80$ at $T= 50$ mK, at different field sweep rates: 0.1 T/hrs (orange), 0.2 T/hrs (blue), 0.5 T/hrs (red) and 1 T/hrs (black), for D2. The curves are vertically shifted by 0.54 $e^2/h$ for clarity. The plateaus are quite robust and the number of plateaus and their step sizes remain nearly unchanged under different magnetic field sweep rates.

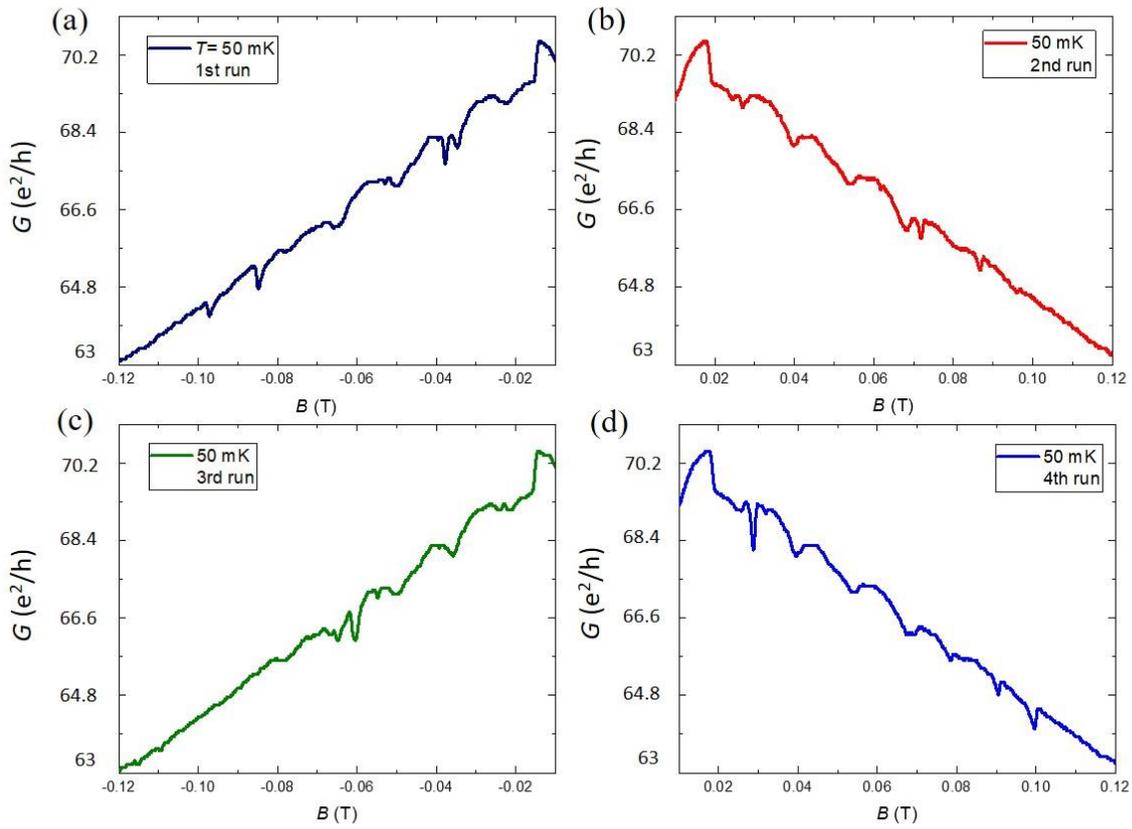

Figure 4. **Reproducibility of the quantized conductance in topological Josephson junctions at different sweep directions:** the quantized conductance as a function of applied out-of-plane magnetic field $B_\perp$ for D3 at 1st (a) and 3rd (c) runs with sweep direction from left to right and, 2nd (b) and 4th (d) runs with sweep direction from right to left. All data are taken at temperature $T$= 50 mK. The quantized plateaus and phase transitions are quite robust and reproducible at different magnetic field sweep directions.

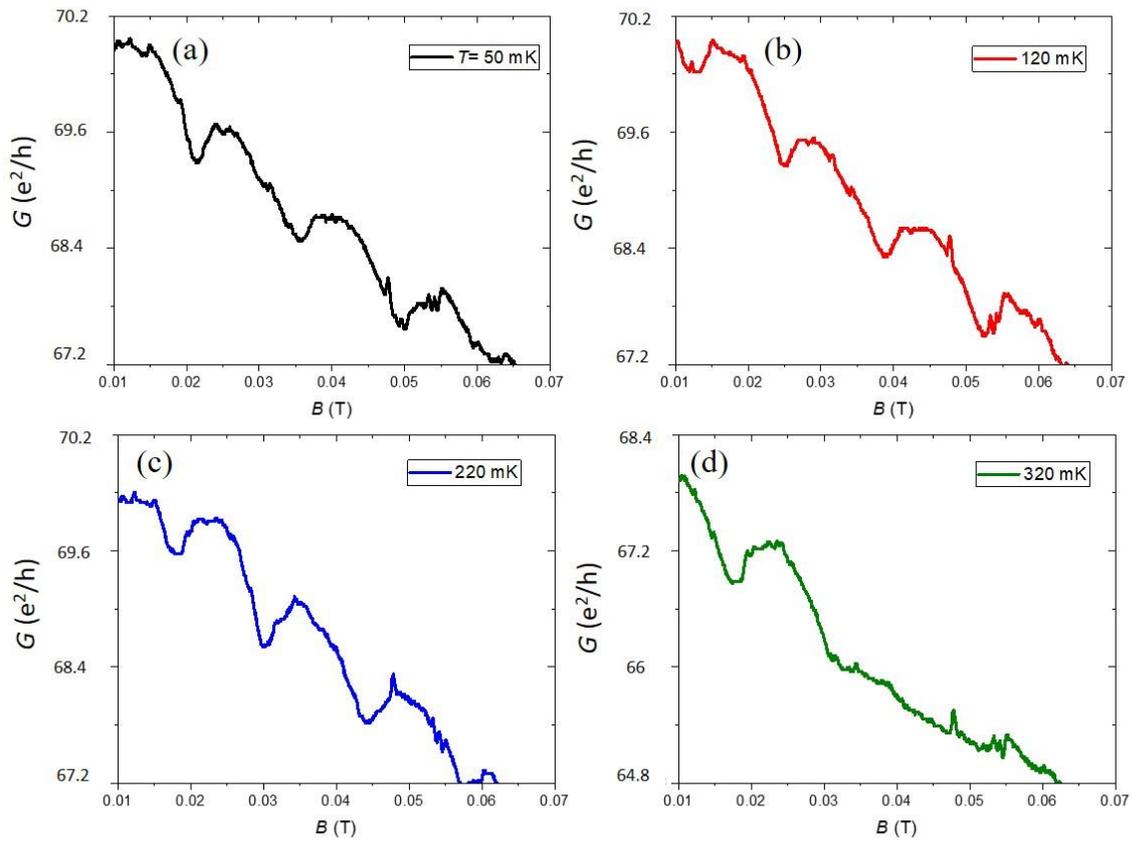

Figure 5. **Temperature dependence of the quantized conductance in topological Josephson junctions:** quantized conductance vs. applied out-of-plane field $B_\perp$ at temperatures $T=$ 50mK (a), $T=$ 120 mK (b), $T=$ 220 mK (c), and $T=$ 320 mK (d), for fields sweep directions from right to left, for D4. The curves are plotted for a selected (small) range of magnetic fields to highlight the robustness of the plateaus and phase transitions.

# SUPPLEMENTARY INFORMATION

# Experimental evidence for topological phases in the magnetoconductance of 2DEG-based hybrid junctions

Kaveh Delfanazari[1,2,3,*], Llorenç Serra[4], Pengcheng Ma[2], Reuben K. Puddy[2], Teng Yi[2], Moda Cao[2], Yilmaz Gul[2], Ian Farrer[2,5], David A. Ritchie[2], Hannah J. Joyce[1], Michael J. Kelly[1,2], Charles G. Smith[2]

[1]Electrical Engineering Division, Engineering Department, University of Cambridge, Cambridge CB3 0FA, UK
[2]Department of Physics, Cavendish Laboratory, University of Cambridge, Cambridge CB3 0HE, UK
[3]James Watt School of Engineering, University of Glasgow, Glasgow G12 8QQ, UK
[4]IFISC (UIB-CSIC) and Physics Department, University of the Balearic Islands, E-07122 Palma, Spain
[5]Department of Electronic and Electrical Engineering, University of Sheffield, Mappin Street, Sheffield, S1 3JD, UK
*Corresponding author email: kd398@cam.ac.uk

Dated: 25062020

Here, we demonstrate the quantum transport in hybrid junctions measured at sub-Kelvin temperature ranges and/or under vertical magnetic fields [1].

Figure **S1** shows the induced superconductivity in $In_{0.75}Ga_{0.25}As$ for four devices (D1-4). All data are taken at $T$= 50 mK. An excess current $I_{exc}$ flows through the junctions as a result of electron- and hole-like quasiparticles correlations (Andreev reflections) below the junctions' transition temperatures and for voltage biases within the superconducting gap. The *dV/dI* ($V_{SD}$) value is reduced in the gap region [2-4], and supercurrents with critical currents of up to 2 µA are measured [1].

Figure **S2** shows the temperature dependence induced superconductivity and supercurrent in $In_{0.75}Ga_{0.25}As$ quantum wells. The *dV/dI* ($V_{SD}$) curves for temperatures between $T$= 50 mK and 800 mK for D4 are plotted in Fig. S2 (a). We observed a flat *dV/dI* ($V_{SD}$) within $\Delta/e$ with subharmonic gap structures (SGS), suggesting highly transparent interfaces between two different materials based on Blonder–Tinkham–Klapwijk (BTK) theory [1,5]. The barrier strength is estimated to be Z < 0.2 (Z is the dimensionless interface barrier strength) in the interfaces between Nb and $In_{0.75}Ga_{0.25}As$, and the induced gap of approximately $100 \leq \Delta_{ind}$ (μeV) $\leq 300$ is measured in D1-4 [1]. The temperature and source-drain voltage dependences induced superconducting gap with pronounced SGS peaks and dips for D4 are shown in Fig. S2 (b). The multiple Andreev reflections (MAR) at the interfaces of the Nb-$In_{0.75}Ga_{0.25}As$-Nb junctions result in the observation of SGS in the differential conductance. At the lowest measured temperature $T=50$ mK, the SGS appears with three peaks (named as P1, P2, and P3 in Fig. S2 (a)) and three dips (named as d1, d2, and d3 in Fig. S2 (a)). The temperature evolution of the peaks and dips due to the suppression of the induced superconductivity with temperature increase can be seen here. The SGS peak positions obey the expression $V = 2\Delta/ne$ ($\Delta$ is the Nb gap energy, $n = 1, 2, 3, \ldots$ is an integer, and $e$ is the electron charge) [1]. All features are significantly temperature-dependent, and the strongest (weakest) SGS peaks (dips) are observed at $T=50$ mK (800 mK). The shape of the gap and SGS has been shown to be influenced by the ratio of $L/\xi_N$ where $\xi_N$ is the Bardeen–Cooper–Schrieffer coherence length [5-7]. The SGS consists of a set of pronounced maxima in $dV/dI$ at $V = 2\Delta/ne$ if $L/\xi_N \ll 1$, but the amplitude of the SGS decreases by increasing the ratio of $L/\xi_N$. For $L \approx \xi_N$ the peaks evolve into dips [7,8]. Both the induced gap and SGS are suppressed significantly at temperatures above $T=400$ mK leading to a shift toward zero bias as shown in Fig. S2 (b). The current–voltage characteristics (*IVC*) at various temperatures is plotted in Fig. S2 (c). A pronounced supercurrent is observed in this high quality Nb-$In_{0.75}Ga_{0.25}As$-Nb junction at zero magnetic fields and low temperatures between $T=50$ and $500$ mK.

The magnetic field dependence of the induced superconductivity is shown in Fig. **S3** for device 4. The color-coded plot of *dV/dI* as a function of applied perpendicular magnetic field $B_\perp$ and applied voltage $V_{SD}$, at $T=50$ mK, is shown in Fig. S3 (a). It can be seen that the induced gap and SGS features that are evidence of enhanced multiple Andreev reflections in high-quality Josephson junctions are quite sensitive to the applied external field. Both effects are suppressed; the position of the peaks shifts toward zero bias, and their amplitudes diminish with further increasing of the magnetic field. Figure S3 (b) shows that the induced superconductivity and SGS are strongly $B_\perp$ dependent, with the position of the peaks bending toward zero voltage as $B_\perp$ is increased. The dark yellow, blue, and dark cyan triangles (see their corresponding arrows in inset) indicate the magnetic field evolution of the induced and SGSs. The plot of $dV/dI$ as a function of applied voltage $V_{SD}$ at $B_\perp = 1$ mT is shown in the inset, demonstrating a robust induced superconducting gap and SGSs in the high-quality Josephson junction.

On our ballistic junctions, the coherent Andreev reflections and correlation of electrons and holes lead to bound states and, therefore, the phase-coherent supercurrents between the electrodes. Figure **S4** (a) shows the *IVC* at the lowest and highest measured temperatures for D3 in the inset. The largest critical current $I_c$ was measured at the lowest temperature of $T = 50$ mK. $I_c$ is suppressed as the temperature increases and disappeared above $T = 350$ mK, as shown by dark cyan triangles. The magnetic field dependence of the induced superconductivity is shown in Fig. S4 (b) and (c) for D3. The applied perpendicular fields affect the phase of the induced superconductivity in the 2DEG that is confined between Nb superconducting electrodes. As a result, the $I_c$ would oscillate, and the period of $I_c$ (B) would change continuously and nonmonotonically from $\Phi_0$ to $2\Phi_0$, where $\Phi_0 = h/2e$ is the flux quantum (Fraunhofer-like) [1,7-9]. However, both trivial [7-9] and topological [10-12] Josephson junctions can reveal widely different oscillation patterns, strongly depending on their topology, dimensions, and material characteristics. For example, $I_c$ in trivial junctions does not decrease to zero if the 2DEG is not enclosed between superconducting contacts [7], or may get a different period due to the change in the effective dimension of junctions [9]. Topological junctions, due to Majorana modes, may show a pattern with a nonzero first minimum, deviating from a standard Fraunhofer pattern [10], or may show a suppression of the odd-numbered lobes in the Fraunhofer pattern [11], or may reveal smaller-than expected flux periods $\Delta B$ [12].

In our junctions, 2DEG is not fully enclosed between two Nb contacts [1,7]. We observe the oscillation of $I_c$, usually with flux periods of around $\Delta B \cong 0.8$ mT [1]. This is approximately corresponding to an area of $A \cong \Delta\Phi/\Delta B \cong 2.5$ µm$^2$ of 2DEG sandwiched between two Nb electrodes which are slightly smaller than the area of the 2DEG between Nb electrodes at the closest path, which should be $\cong 3$ µm$^2$ based on the CAD design [1]. In the presence of the vertical magnetic field, we, therefore, estimate that effective dimension of the junctions are slightly changed [1,7-12] and their width $w$ and length $L$ may get the sizes between $0.8 \leq w$ (µm) $\leq 2$ and $2 \leq L$ (µm) $\leq 1$ taking in to account the coherence length of $\approx 2$ µm in 2DEG in our ballistic and highly coupled junctions [1]. This would result in the number of transport modes between $60 \leq N \leq 80$ in the conductance of the junctions [13] and is in a good agreement with the conductance of the junctions calculated based on our model and also measured in the experiment.

Figures **S5** and **S6** are the theoretical modeling based on the approach of [14] for a narrower ($w$= 0.75 µm) and wider ($w$= 1.25 µm) Josephson junction, respectively, as compared to the 1 µm wide junction shown in the main text (Fig.1 b-c). It is clearly seen how for wider junctions, the topological phases are more densely packed in the µ-$B$ plane and how this leads to a sequence of very closely spaced conductance dips and plateaus as a function of the magnetic field. We note that already for $w$= 1.25 µm resolving the sequence of such transitions with increasing fields requires an extremely fine magnetic sweep, at the limit of the numerical possibility and well beyond the experimental one.

Figure **S7** (a) shows a phenomenological model for a hysteretic $B_{\text{eff}}$(B) similar to Ref [15], although with *much smaller* internal field enhancements. We assume that the switching field $B_s$ is very close to zero, and it signals the position of a sudden jump in $B_{\text{eff}}$ due to the reorientation of the internal magnetization. Figure **S7** (b) shows that the phenomenological magnetoconductance Equation (4), see Materials and Methods, when presented as $G(B_{\text{eff}}(B))$, qualitatively reproduces the experimental hysteresis between forward and backward sweeps of the magnetic field. This phenomenological model, including very small internal magnetic interactions, agrees with the experiment about the presence of a sequence of conductance dips at a rather low field as well as the presence of the hysteresis.

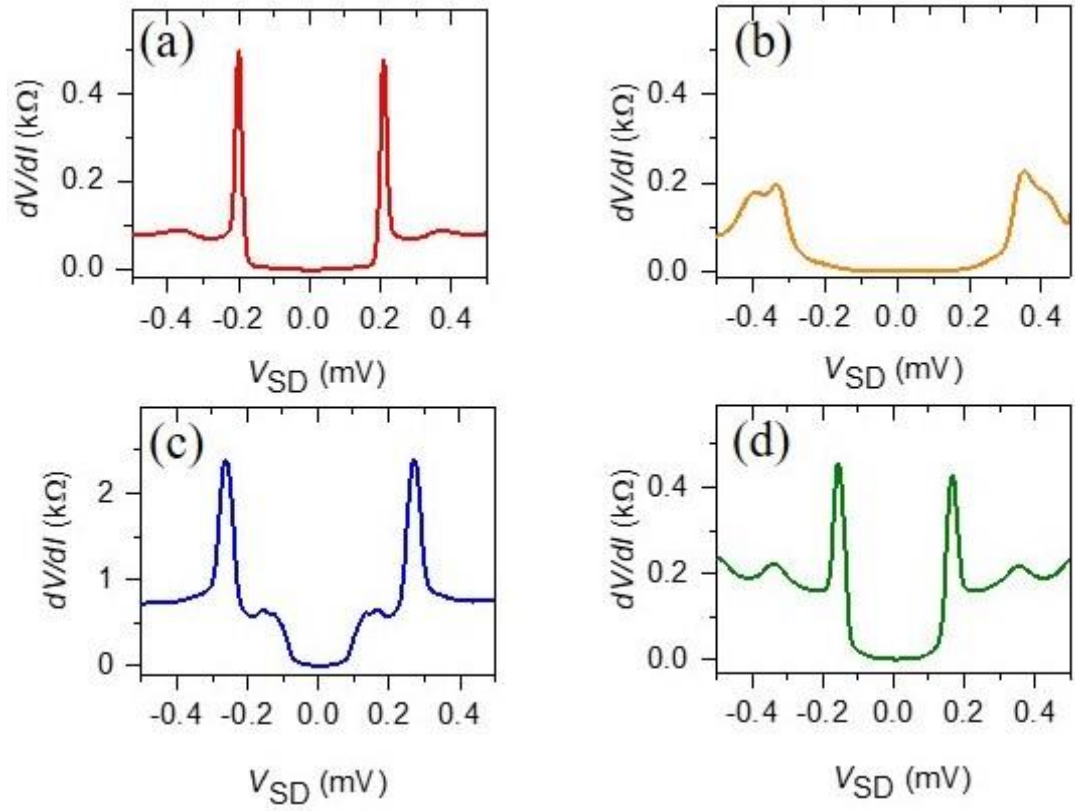

**Figure S1.** Induced superconducting properties in In$_{0.75}$Ga$_{0.25}$As quantum wells in *D*1-4. All curves are for base temperature *T*= 50 mK.

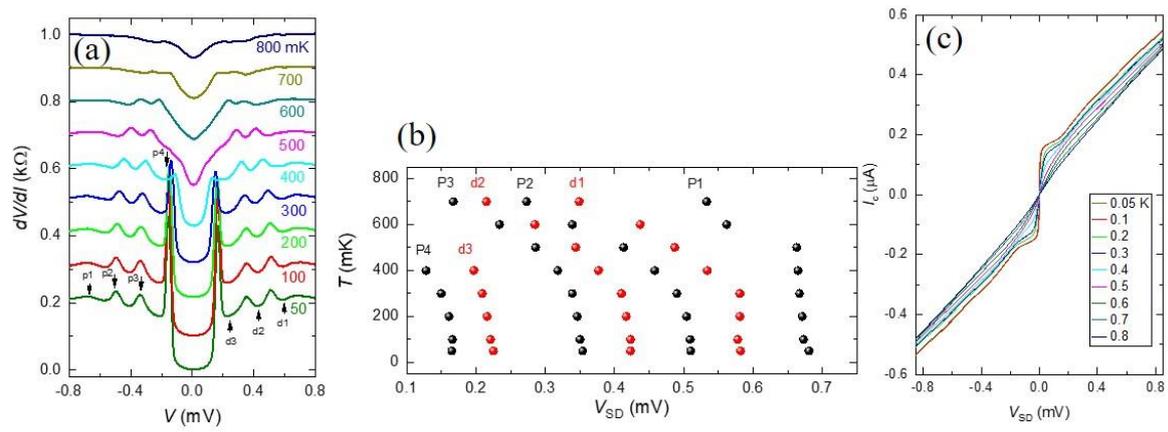

**Figure S2.** Temperature dependence induced superconducting gap with subharmonic-gap structures (SGS) oscillations due to multiple Andreev reflections observed in *D*4. The SGS and the induced gap peaks, are denoted by *P* while the dips are marked by *d*. (b) The SGS peaks and dips shown in (a) as a function of temperature and $V_{SD}$. (c) The current–voltage characteristics (*IVC*) curves as a function of temperature. The supercurrent in the Josephson junction is quite sensitive to temperature.

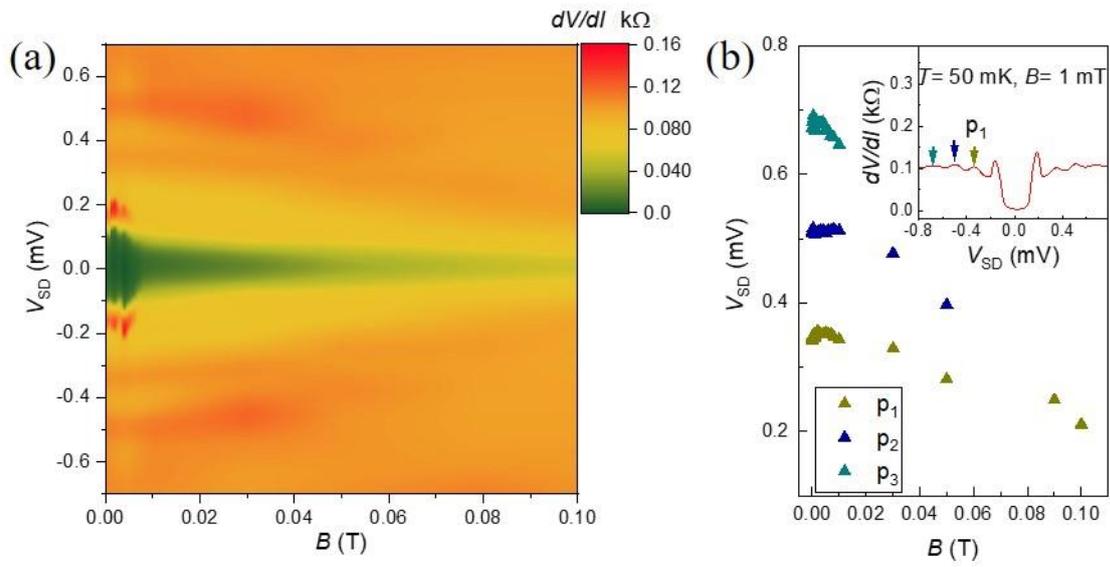

**Figure S3.** (a) Magnetic field dependence of induced superconductivity as a function of $V_{SD}$ at temperature $T$= 50 mK, in $D$4. (b) The dark yellow, blue, and dark cyan triangles (see their corresponding arrows in inset) indicate the magnetic field evolution of the induced- and SGS. Inset is the $dV/dI$ (V) measured at $B_⊥$= 1 mT and $T$= 50 mK. Arrows correspond to SGS peaks.

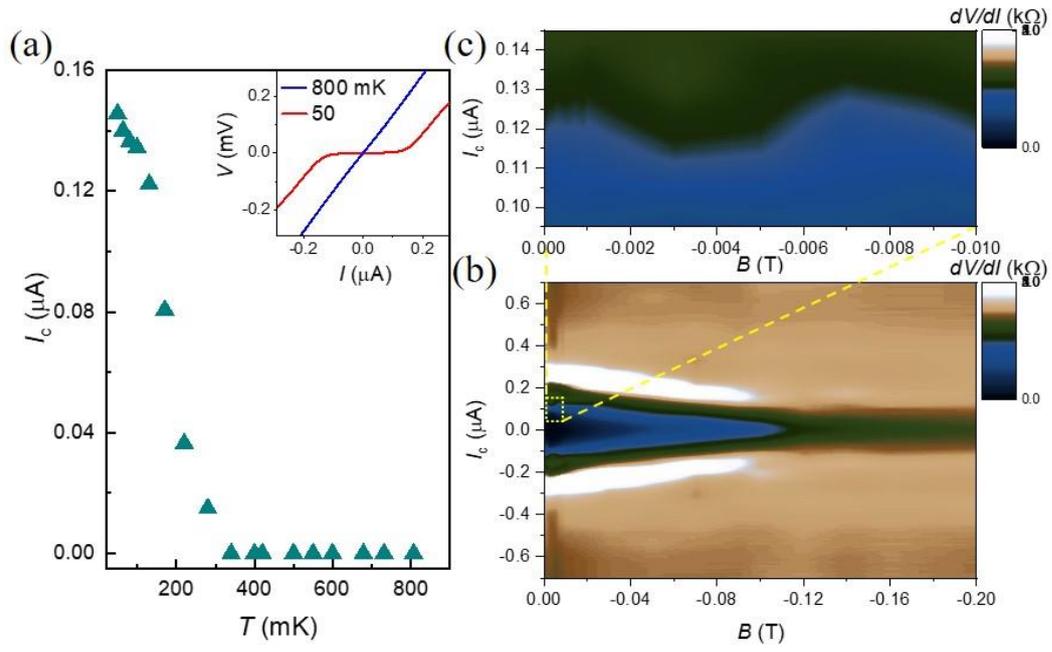

**Figure S4.** (a) The current–voltage characteristics (*IVC*) at *T*= 50 mK and 800 mK (inset) and the critical current $I_c$ as a function of temperature for *D*3. (b) 2D plots of *dV/dI* as a function of $I_c$ and $B_\perp$, at *T*= 50 mK. (c) 2D plots of *dV/dI* as a function of $I_c$ and $B_\perp$ shown for smaller magnetic field regions to highlight the supercurrent oscillation in the junction under application of the out-of-plane magnetic fields $B_\perp$.

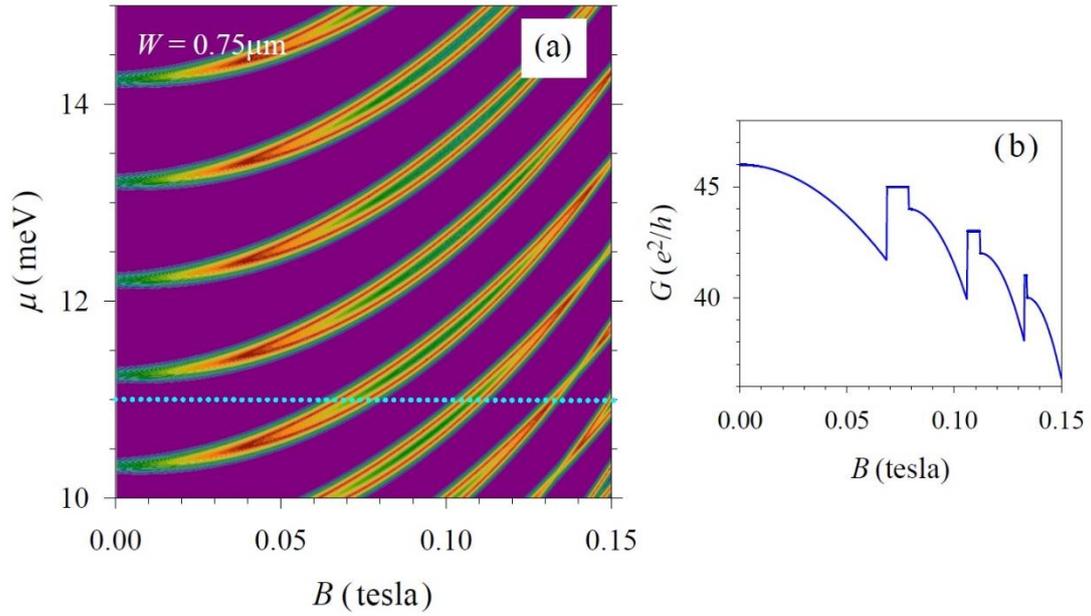

**Figure S5.** (a) Phase diagram of the hybrid 2DEG junction similar to Fig. 1b of the main text. The junction width is $w$= 0.75 μm and the induced superconductivity is $\Delta_0$= 60 μeV. (b) Theoretical conductance of the junction in the phenomenological model of our work along the sequence of magnetic fields of the line cut in panel (a). As explained in the main text, the model assumes a perfect Andreev reflection in the topological phases (plateaus), and a decreasing Andreev reflection AR (dips) with increasing field in the trivial phases due to interface scattering.

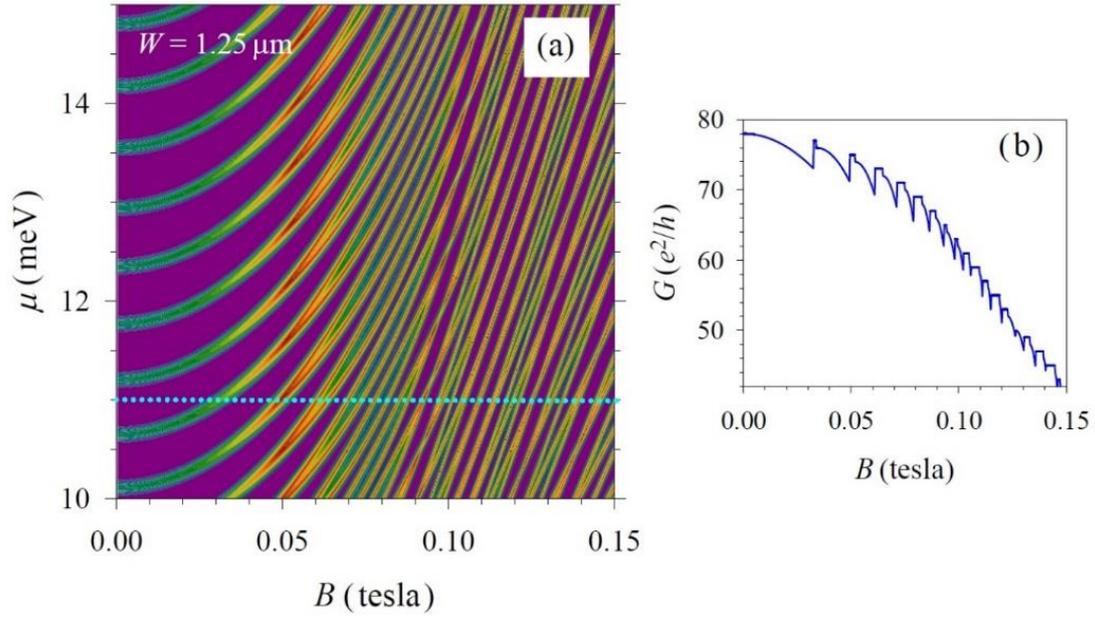

**Figure S6.** (a) Phase diagram of the hybrid 2DEG junction similar to Fig. 1b of the main text. The junction width is $w$= 1.25 μm and the induced superconductivity is $\Delta_0$= 60 μeV. (b) Theoretical conductance of the junction in the phenomenological model of our work along the sequence of magnetic fields of the line cut in panel (a). As explained in the main text, the model assumes a perfect Andreev reflection in the topological phases (plateaus), and a decreasing Andreev reflection $AR$ (dips) with increasing field in the trivial phases due to interface scattering.

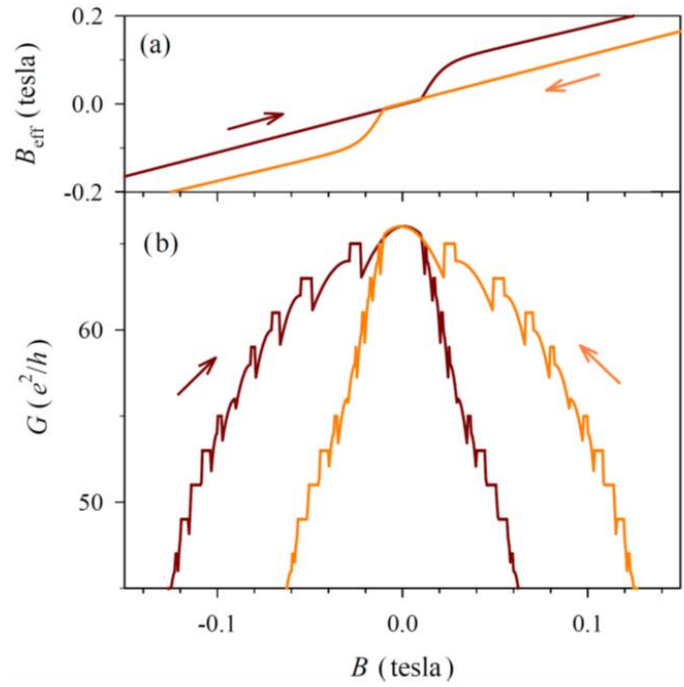

**Figure S7.** Numerical simulation of the magnetoconductance in topological Josephson junctions: (a) model of the hysteresis loop of the internal effective field $B_{eff}$ when sweeping the external field $B$. The arrows indicate the sweep direction. (b) the modeled quantized conductance vs. applied out-of-plane magnetic field $B_\perp$ in hybrid Josephson junction, explained in the text. The directions of external field $B_\perp$ sweeps indicated by the arrows.